\documentclass[]{interact}

\usepackage{epstopdf}
\usepackage{subfigure}
\usepackage{algorithm}
\usepackage{algorithmic}
\usepackage{float}

\usepackage{natbib}
\bibpunct[, ]{(}{)}{;}{a}{}{,}

\theoremstyle{plain}

\theoremstyle{definition}

\theoremstyle{remark}

\author{
Kamil Bujel, Feiko Lai, Michal Szczecinski, Winnie So, Miguel Fernandez\\
Dept. of Analytics, GoGo Tech Limited, Hong Kong
}

\title{Solving High Volume Capacitated Vehicle Routing Problem with Time Windows using Recursive-DBSCAN clustering algorithm}

\begin{document}
\maketitle

\begin{abstract}
This paper introduces a new approach to improve the performance of the Capacitated Vehicle Routing Problem with Time Windows (CVRPTW) solvers for a high number of nodes. It proposes to cluster nodes together using Recursive-DBSCAN - an algorithm that recursively applies DBSCAN until clusters below the preset maximum number of nodes are obtained. That approach leads to 61\% decrease in runtimes of the CVRPTW solver as benchmarked against Google Optimization Tools, while the difference of total distance and number of vehicles used by found solutions is below 7\%.  The improvement of runtimes with the Recursive-DBSCAN method is because of splitting the node-set into constituent clusters, which limits the number of solutions checked by the solver, consequently reducing the runtime. The proposed method consumes less memory and is able to find solutions for problems up to 5000 nodes, while the baseline Google Optimisation Tools solves problems up to 2000 nodes.
\end{abstract}


\section{Introduction}

In the current sharing and `on demand' economy, the number of services available and the demand for them have increased \citep{Schor15}. New services, such as Uber, DiDi or GOGOVAN have emerged. One of the benefits of such a system is the ability to develop optimisation logic that increases profits for partners, and decreases the costs for customers \citep{Nguyen18}. It is no different in logistics, where solving routing problems makes goods delivery more efficient by optimising fleets of vehicles - allowing them to deliver more parcels and helping each driver to complete more deliveries on a daily basis.

Problems such as the Travelling Salesperson Problem (TSP) or Vehicle Routing Problem (VRP) have always been given much attention in the literature. TSP was first explored by \citep{Flood56} and VRP was introduced using the example of optimal routing of gasoline trucks by \citep{Dantzig59}. Ever since, both of these problems have been important ones in the field of Operations Research.

In this work, we focus on solving the special case of the Vehicle Routing Problem, where customers must be served within the given time window and each of the vehicles has a limited capacity - namely the Capacitated Vehicle Routing Problem with Time Windows (CVRPTW). Vehicle Routing Problem with Time Windows was proved to be NP-hard \citep{Solomon86}.

Throughout the time, there have been many proposed solutions to the CVRPTW. The first exact method to be proposed was the branch-and-bound approach\citep{Christofides69}, which is the basis for the branch-and-cut exact algorithm \citep{Baldacci04}. However, even the current state-of-the-art exact methods work only for up to 360 waypoints \citep{Pecin17}, which is not enough in our use case.

Another approach is to use metaheuristics in order to improve a suboptimal solution. Three metaheuristics \citep{Laporte06} are the most efficient: Genetic Algorithms (GA) \citep{Baker03} \citep{Alba06} \citep{Karakatic15}, Local Search \citep{Cordeau98}, \citep{Nagata08} or Ant Colony Optimisation (ACO) \citep{Gambardella99}. 

The most current benchmarks indicate that the method by \citep{Pecin17} performs the fastest, as tested for up to 1000 nodes \citep{Uchoa16}. 

There have been quite a few frameworks that are based on these algorithms, both open source and commercial. For open source solutions, there are VRPH \citep{Groer10} and Google Optimization Tools (OR-tools) \citep{Perron11}. While for commercial solutions, Gurobi \citep{Gurobi18}, LocalSolver \citep{Zhang15} and other Constraint Programming solvers are used.

In this paper, we focus on improving the performance of the OR-tools framework by using a modified version of the DBSCAN clustering algorithm \citep{Ester96}, which we refer to as "Recursive-DBSCAN". The reason why we decided to do so, is because the runtimes of OR-tools solvers increase rapidly (Fig. 1) and also, consume a lot of RAM memory for problems with more than 1000 waypoints. Locally, we managed to run it for a maximum of 2000 waypoints, which consumed 20GB of memory.

\begin{figure}[H]
\centering
\resizebox*{10cm}{!}{\includegraphics{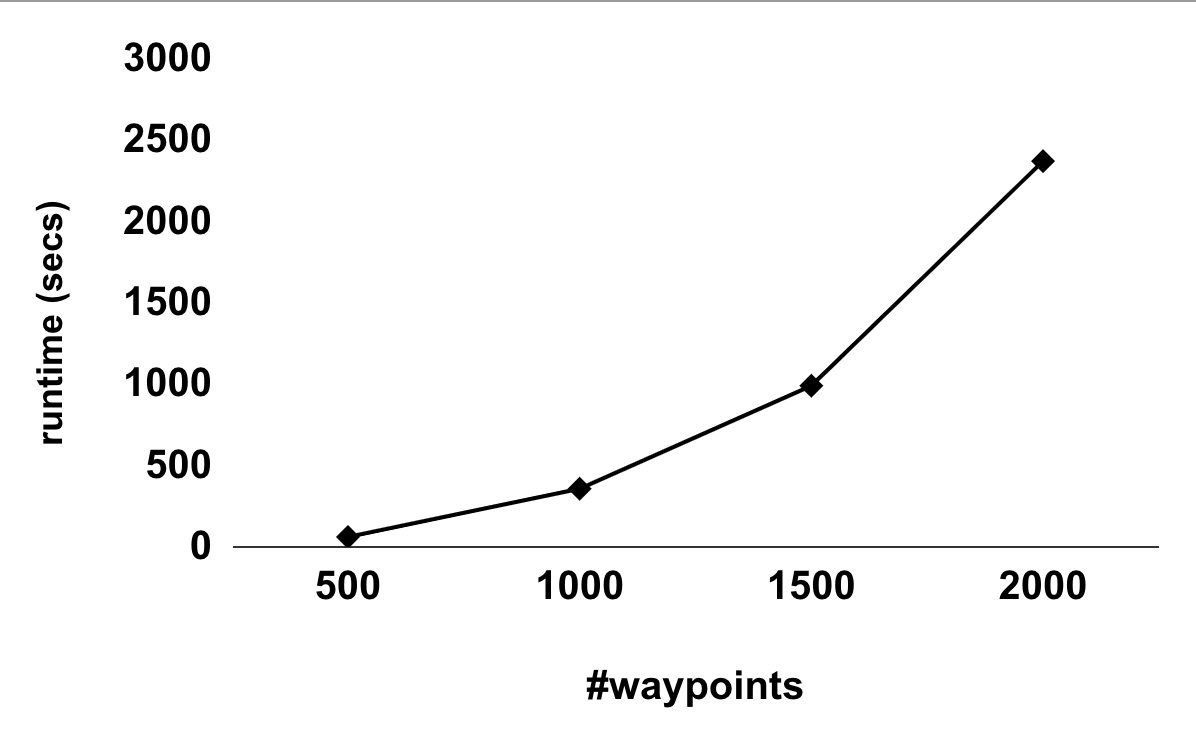}}
\caption{Runtime of CVRPTW solver using OR-tools.} \label{or_tools_runtime}
\end{figure}

In the Section 2 of this paper we formally formulate our problem. Section 3 describes our proposed method, with the results being presented in Section 4 and discussed in Section 5. Section 6 proposes some future work on the problem.

\section{Problem Formulation}

Vehicle Routing Problem can be subjected to various constraints such as constraints on vehicle capacity and time window of arrival. We look into VRP's particular case - Capacitated Vehicle Routing Problem with Time Windows. We can describe the problem as follows:

Goods are to be delivered to a set of delivery points by a number of vehicles departing from the central depot.  The vehicles are supposed to pick up all the packages from the depot within a certain time window. Similarly, each  delivery vehicle needs to arrive at each delivery point within a certain time window specific to that point. Each vehicle has a limited capacity and each delivery has a certain size.

The goal is to determine a route schedule which minimises the travelled distance, given the following constraints:

\begin{enumerate}
	\item Each vehicle starts its route at the depot.
	\item The total size of packages assigned to a particular driver cannot exceed the capacity of the vehicle.
	\item Each delivery waypoint is to be served by one and only one vehicle.
	\item Each delivery needs to be completed within a given time window.
\end{enumerate}

There are N delivery waypoints to be visited by the maximum of M vehicles. Let's assume that the depot is node $0$. The size of each delivery $i$, $i \in [1, 2, 3, ..., N]$, is defined as $q_i$, while the time window is $[e_i, l_i]$. The pickup time at the depot is $[e_0, l_0]$. Each vehicle $k$, $k \in [1, 2, 3, ..., M]$, has got a maximum capacity $Q_k$. We define travel time between nodes $i$ and $j$ for the vehicle $k$ as $t_{i,j}^k$,  while $d_{i,j}^k$ is the distance between nodes $i$ and $j$ for the vehicle $k$.

As in our case, we do not force vehicles to return back to the depot, we set $t_{i, 0}^k=0$

We can describe the problem mathematically \citep{Szeto11} as:
\begin{equation}
\text{Minimise } \sum^M_{k=1}\sum^N_{i=0} \sum^N_{j=0}{d_{i,j}^k x_{i,j}^k}
\end{equation}
Subject to:
\begin{equation}
x_{i,j}^k = 
\begin{cases}
  1$, if vehicle $k$ travels from $i$ to $j.\\
  0$, otherwise$
\end{cases}
\end{equation}
\begin{equation}
\sum_{k=1}^M \sum_{i=0}^N{x_{i,j}^k}=1 \text{ for } j \in [1, 2, 3, ..., N]
\end{equation}
\begin{equation}
\sum_{k=1}^M \sum_{j=0}^N{x_{i,j}^k}=1 \text{ for } i \in [1, 2, 3, ..., N]
\end{equation}
\begin{equation}
\sum_{i=0}^N{x_{i,l}^k} - \sum_{j=0}^N{x_{l,j}^k} = 0 \text{ for } k \in [1, 2, 3, ..., M], l \in [1, 2, 3, ..., N]
\end{equation}
\begin{equation}
\sum^N_{j=0}{q_j}(\sum_{i=0}^N{x_{i,j}^k}) \leq Q_k \text{ for } k \in [1, 2, 3, ..., N]
\end{equation}
\begin{equation}
e_i \leq a_i \leq l_i \text{ for } i \in [1, 2, 3, ..., N]
\end{equation}
\begin{equation}
e_0 \leq p_k \leq l_0 \text{ for } k \in [1, 2, 3, ..., M]
\end{equation}
\begin{equation}
\sum_{j=1}^N{x_{0,j}^k} \leq 1 \text{ for } k \in [1, 2, 3, ..., M]
\end{equation}
\begin{equation}
\sum_{i=1}^N{x_{i,0}^k} \leq 1 \text{ for } k \in [1, 2, 3, ..., M]
\end{equation}

where $N$ is the number of deliveries, $M$ is the number of delivery vehicles, $a_i$ is the arrival time of the vehicle at node $i$ and $p_k$ is the arrival of the vehicle $k$ at the depot.

The objective function in Eq. (1) is to minimise the total travelled distance for all the vehicles. Constraints in Eq. (3) and Eq. (4) ensure that one and only one vehicle respectively arrives at and departs from each of the nodes. Eq. (5) ensures the route continuity. Eq. (6) ensures that the total size of packages taken by a vehicle does not exceed its capacity. Eq. (7) and Eq. (8) ensure that the time windows are met. Eq. (9) and Eq. (10) make sure that each vehicle is used no more than once.
\section{Method}

\subsection{Data Collection and Processing}

The datasets are prepared from a sample of GOGOVAN trips completed in Hong Kong in 2017. They contain more than 10,000 delivery waypoints, provided in the GPS format. The datasets are randomised during the analysis.

The tests are conducted on a 2017 Macbook Pro 13", with 2.3 GHz dual-core Intel Core i5, 16GB 2133 Mhz LPDDR3 RAM and 256GB SSD. For distance calculation, we use straight-line distance.

\subsection{Algorithms for Vehicle Routing Problem}

We decided to present our proposed Recursive-DBSCAN solution and benchmark it against two other methods.

The first method involves passing all the waypoints and vehicles, along with the constraints to the Google Optimization Tools solver and treating its output as the final solution. The other two methods are clustering-based approaches, where we first attempt to cluster the delivery waypoints and then try to optimise each of the clusters separately, by also passing it to the OR-tools Routing solver. The second method uses classical DBSCAN, while the third one is the proposed Recursive-DBSCAN clustering approach.

\subsubsection{Google Optimization Tools}
As a baseline, we chose to solve the CVRPTW problem using the OR-tools library.  

The parameters we call Google Optimization Tool's Routing library with every time are as follows:

\begin{table}[H]
\caption{Google Optimisation Tools Search Parameters.}
\centering 
\hspace*{-1cm}
\begin{tabular}{lc p{8cm}} \toprule
 Parameter & Value & Description \\ \midrule
  first\_solution\_strategy & PATH\_CHEAPEST\_ARC & First solution strategies, used as starting point of local search. 	Starting from a route "start" node, connect it to the node which produces the cheapest route segment, then extend the route by iterating on the last node added to the route. \\ \\
 optimization\_step & 1 & Minimum step by which the solution must be improved in local search. \\ \\
 solution\_limit & 9223372036854775807 & Limit to the number of solutions generated during the search. \\ \\
 time\_limit\_ms & 5000 & Limit in milliseconds to the time spent in the search.  \\ \\
 use\_light\_propagation & true & 	Use constraints with light propagation in routing model. 
 Extra propagation is only necessary when using depth-first search or for models which require strong propagation to finalise the value of secondary variables. Changing this setting to true will slow down the search in most cases and increase memory consumption in all cases. \\
 \bottomrule
\end{tabular}
\hspace*{-1cm}
\label{abc}
\end{table}

\subsubsection{DBSCAN}
One of the approaches explored is to use the DBSCAN clustering algorithm to cluster the waypoints to be optimised into several clusters of different size that will then be optimised one by one using the OR-tools method.

We define one very important constant here - $max\_cluster\_size\_const$, which represents the maximum number of delivery points in each cluster. Our goal is to obtain the clusters of maximum possible average size, while each of them is smaller than the aforementioned constant. We then pass these clusters as input to the OR-tools routing solver to obtain the optimised routes.

The pseudocode of such a method looks as follows:

\begin{algorithm}[H]
\caption{CVRPTW with DBSCAN}
\begin{algorithmic}
\REQUIRE $(params, min\_radius, max\_radius, max\_cluster\_size\_const) = const$ 
\REQUIRE $\text{waypoints} = [1, 2, ..., N] \text{ , } \text{vehicles} = [1, 2, ..., M]$
\STATE $\text{m\_per\_radian} = 6,371,008.8 \text{ m rad}^{-1}$
\STATE $best\_avg\_cluster\_size = 0$
\WHILE {$min\_radius < max\_radius$}
\STATE $radius \leftarrow (min\_radius+max\_radius)/2$
\STATE $\epsilon \leftarrow radius / m\_per\_radian$
\STATE $clusters \leftarrow DBSCAN(orders, \epsilon, algorithm='ball\_tree', metric='haversine')$
\IF{$max\_cluster\_size > max\_cluster\_size\_const$}
\STATE $max\_radius \leftarrow radius - 1$
\ELSE
\STATE $min\_radius \leftarrow radius + 1$
\IF {$avg\_cluster\_size > best\_avg\_cluster\_size$}
\STATE $best\_avg\_cluster\_size \leftarrow avg\_cluster\_size$
\STATE $final\_clusters \leftarrow clusters$
\ENDIF
\ENDIF
\ENDWHILE
\IF{$final\_clusters == \varnothing$}
\STATE return $NO\_SOLUTION\_FOUND$
\ENDIF
\STATE $optimised\_routes = ortools\_optimise(final\_clusters)$
\STATE return $optimised\_routes\_clusters$
\end{algorithmic}
\end{algorithm}

Such a method already divides the routing problem into several smaller problems, consequently improving the performance of the algorithm.

\subsubsection{Recursive-DBSCAN}

We also propose a modification of the method described above, which applies DBSCAN recursively to each cluster, with the aim to make clusters more well-balanced in terms of number of waypoints in each of them (while decreasing the spatial balance). 

We use binary search to find the initial feasible radius that maximises $avg\_cluster\_size$. We then identify clusters which contain too many nodes and apply recursively the same binary search logic in order to break down those large clusters into smaller, constituent clusters. Such an approach ensures that the found final clusters are of maximum possible average size, while at the same time they all lie within the preset size boundaries.

Such a method creates clusters of large radii in the less dense regions and smaller clusters in the high-density regions. The aim of that method is to obtain clusters that contain a similar number of waypoints and yet different radii. Such an approach resolves one issue encountered with classical DBSCAN - clusters in regions with more nodes contain more nodes than clusters in less dense regions. It also avoids a situation when a vehicle is assigned to a point within a cluster that contains only a few nodes.

We have a few constraints, namely: $max\_cluster\_size\_const$, $min\_cluster\_size\_const$ and $min\_no\_clusters$.

By reducing the variance of the number of nodes in clusters and putting constraints on both the maximum and minimum number of nodes per cluster, we obtain splits which can be effectively optimised separately, as they are smaller in terms of number of nodes. That should lead to improvement in runtimes of classical VRP solvers.

The pseudocode of the proposed can be found below:

\begin{algorithm}[H]
\caption{Recursive-DBSCAN}
\begin{algorithmic}
\REQUIRE $(params, min\_radius, max\_radius, max\_cluster\_size\_const = const$
\STATE $min\_cluster\_size\_cons, min\_no\_clusters) = const$ 
\REQUIRE $\text{waypoints} = [1, 2, ..., N] \text{ , } \text{vehicles} = [1, 2, ..., M]$
\STATE $\text{m\_per\_radian} = 6,371,008.8 \text{ m rad}^{-1}$
\STATE $best\_avg\_cluster\_size = 0$
\WHILE {$min\_radius < max\_radius$}
\STATE $radius \leftarrow (min\_radius+max\_radius)/2$
\STATE $\epsilon \leftarrow radius / m\_per\_radian$
\STATE $clusters \leftarrow DBSCAN(orders, \epsilon, algorithm='ball\_tree', metric='haversine')$
\IF{$no\_clusters < min\_no\_clusters$} 
\STATE $max\_radius \leftarrow radius - 1$
\ELSE
\STATE $min\_radius \leftarrow radius + 1$
\IF {$avg\_cluster\_size > best\_avg\_cluster\_size$}
\STATE $best\_avg\_cluster\_size \leftarrow avg\_cluster\_size$
\STATE $best\_clusters \leftarrow clusters$
\ENDIF
\ENDIF
\ENDWHILE
\IF{$best\_clusters == \varnothing$)}
\STATE return $NO\_SOLUTION\_FOUND$
\ENDIF
\STATE $final\_clusters \leftarrow \varnothing$
\FOR{$cluster \in best\_clusters$}
\IF{$len(cluster) > max\_cluster\_size\_const$}
\STATE $new\_clusters \leftarrow recursive-dbscan(cluster, best\_radius)$
\STATE $final\_clusters \leftarrow final\_arrow \cup new\_clusters$
\ELSE
\STATE $final\_clusters \leftarrow final\_arrow \cup cluster$
\ENDIF
\ENDFOR
\STATE $optimised\_routes = ortools\_optimise\_clusters(final\_clusters)$
\STATE return $optimised\_routes$
\end{algorithmic}
\end{algorithm}

\subsection{Optimisation with OR-tools call}

Having already produced the clusters using either the DBSCAN or the Recursive-DBSCAN method, we  have obtained clusters of delivery points ready to be optimised. We attempt to optimise cluster-by-cluster and with each iteration we remove vehicles marked as ``busy'' - which means these ``busy'' vehicles are not considered for assignment for other clusters.

The following pseudocode illustrates that method:

\begin{algorithm}[H]
\caption{ortools\_optimise\_clusters()}
\begin{algorithmic}
\FOR{$cluster \in final\_clusters$}
\IF{$vehicles == \varnothing$}
\STATE return $NO\_SOLUTION\_FOUND$
\ENDIF
\STATE $result\_route, vehicles\_busy \leftarrow ORTOOLS\_OPTIMISE(cluster, vehicles, params)$
\STATE $routes \leftarrow routes  \cup  result$
\STATE $vehicles \leftarrow vehicles \setminus vehicles\_busy $
\ENDFOR
\STATE return $routes$
\end{algorithmic}
\end{algorithm}

\subsection{Setting of Testing Parameters}

\begin{figure}[H]
\centering
\resizebox*{9cm}{!}{\includegraphics{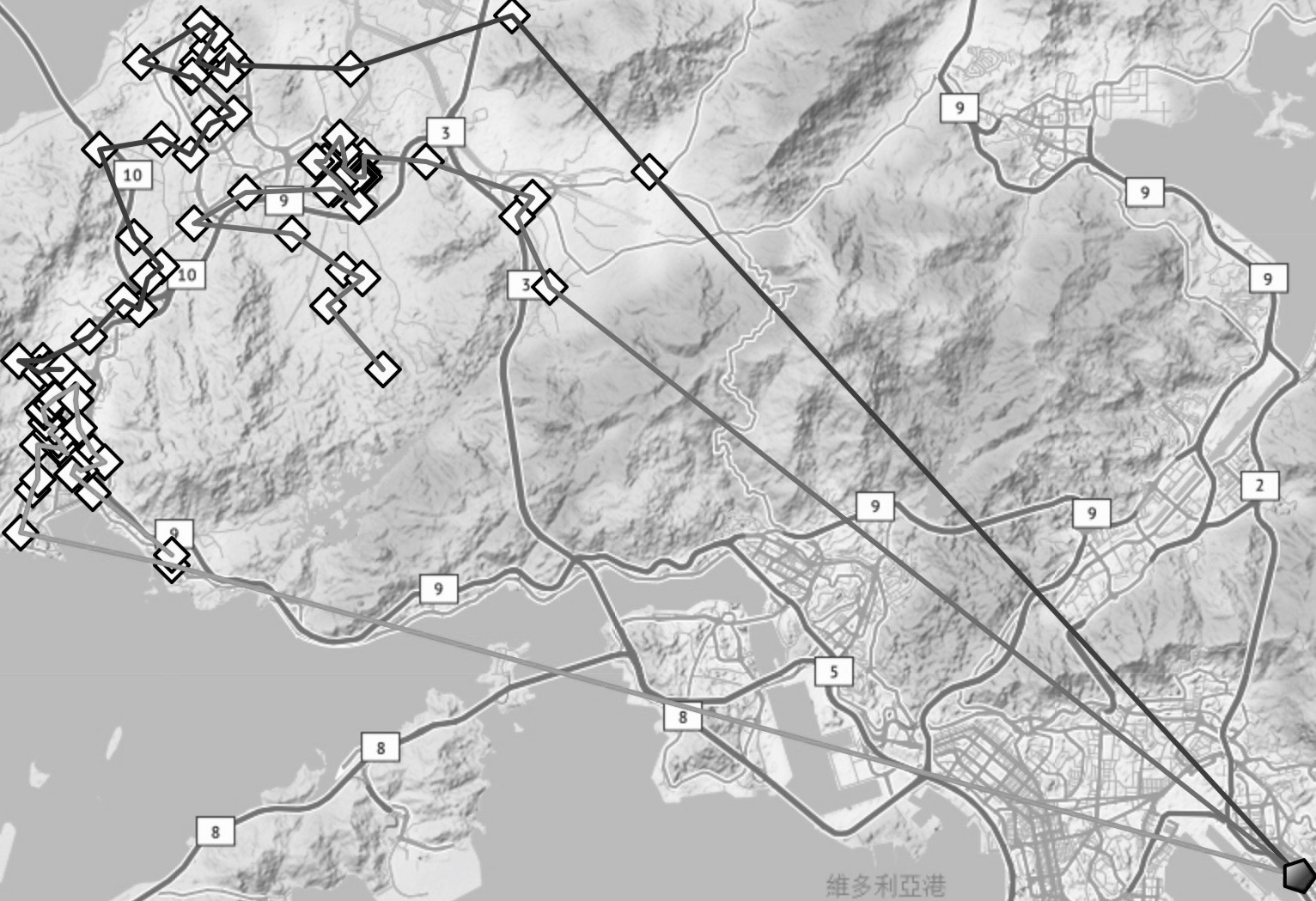}}
\caption{A sample of optimised routes using the Recursive-DBSCAN method for Hong Kong.} \label{results-map-recursive}
\end{figure}

The algorithm's final output is an optimised delivery route, a sample of which is visualised in Fig. 2. 

To test the performance. we chose the batches of size between 500 and 5000 waypoints, every 500. Three different metrics were used to measure the performance: runtime, total distance of the routes and the number of vehicles that are busy.  For each data point, there were 15 different tests conducted.

Also, the following parameters presented in Table 2 were used during the analysis:

\begin{table}[H]
\caption{Testing parameters}
\centering 
\begin{tabular}{lcc} \toprule
 Parameter & Value \\ \midrule
  capacity of each vehicle & 30  \\
  max\_cluster\_size & 500  \\
 min\_cluster\_size & 35  \\ 
 min\_radius & 1 \\
 max\_radius & 10000 \\
  \bottomrule
\end{tabular}
\label{abc}
\end{table}

Our goal in this test was to reduce the runtime, without much impact on total distance and the number of vehicles. The OR-tools result, as presented in Fig. 1., was treated as the benchmark.

\section{Results}

\subsection{Total Travelled Distance}

\begin{figure}[H]
\centering
\resizebox*{9cm}{!}{\includegraphics{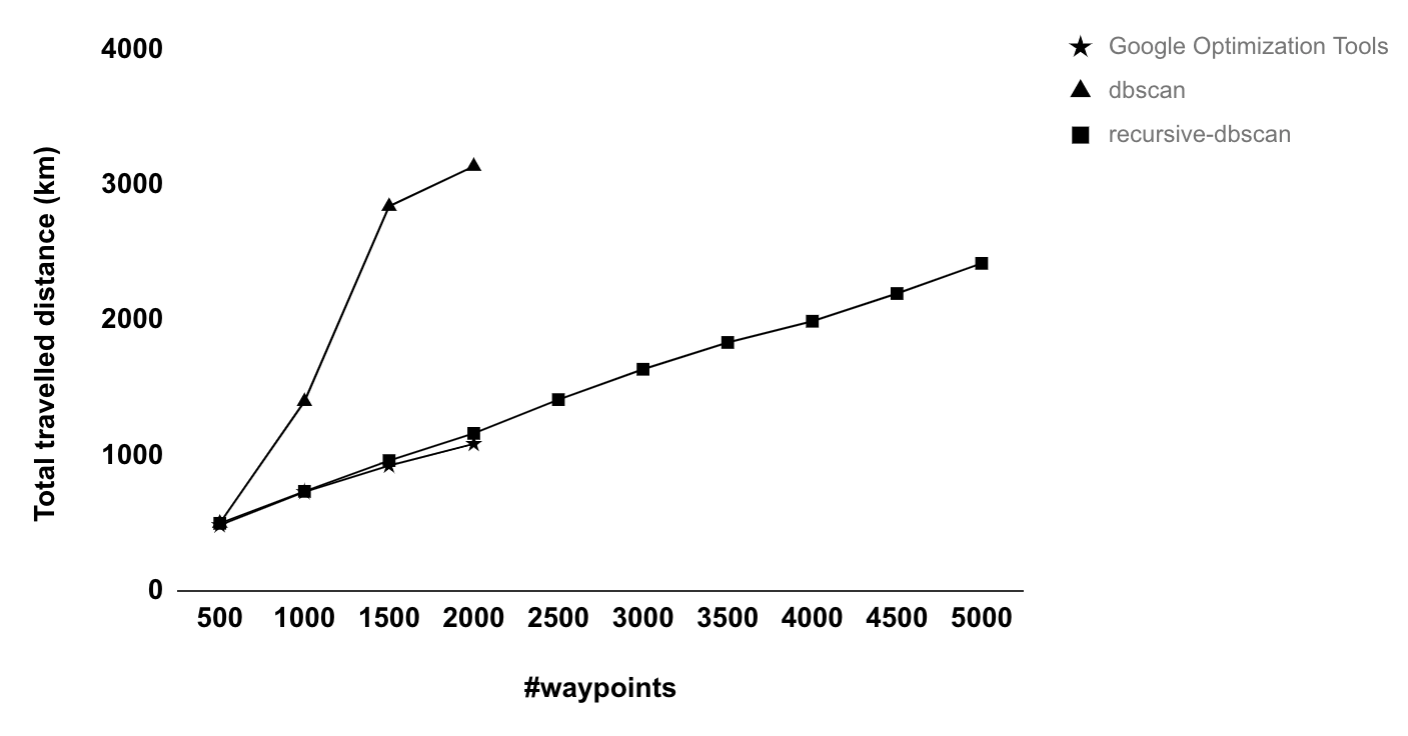}}
\caption{Average total travelled distance by all vehicles for each solution.} \label{results-travelled-distance}
\end{figure}

As can be seen in Fig. 3, for DBSCAN and Recursive-DBSCAN the distance increases linearly with the waypoints. The gradient of the Recursive-DBSCAN curve is similar to the gradient of the OR-tools. DBSCAN curve increases more rapidly than both Recursive-DBSCAN and OR-tools curves.

The DBSCAN curve is cut off at 2000 waypoints due to consuming more than 20GB of memory. That caused the program to crash.

\subsection{Average Number of Busy Vehicles}

\begin{figure}[H]
\centering
\resizebox*{9cm}{!}{\includegraphics{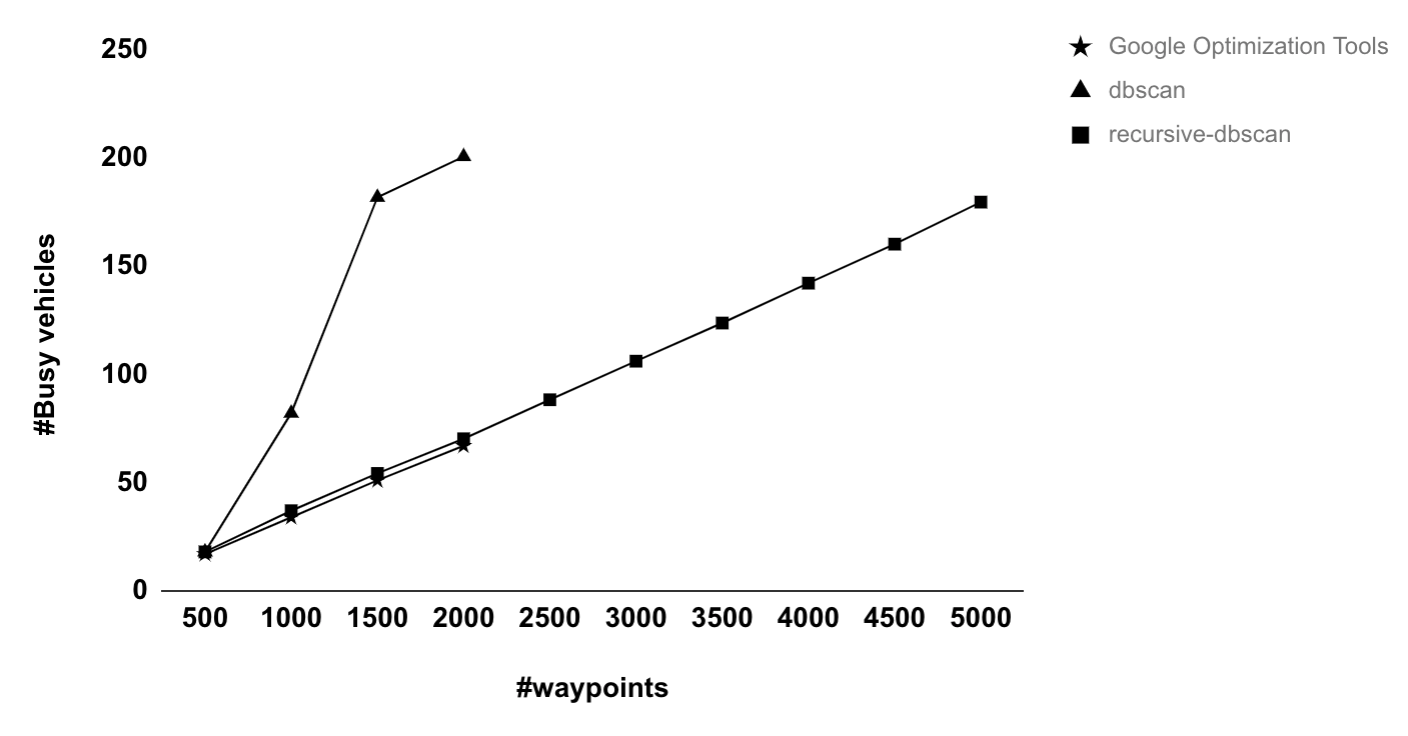}}
\caption{Average \#busy vehicles for each solution.} \label{results-vehicles}
\end{figure}

Similarly to the Total Travelled Distance, Fig. 4. shows that OR-tools and Recursive-DBSCAN increase linearly with the number of waypoints, while DBSCAN solution has got a sharp increase.

\subsection{Runtime}

\begin{figure}[H]
\centering
\resizebox*{9cm}{!}{\includegraphics{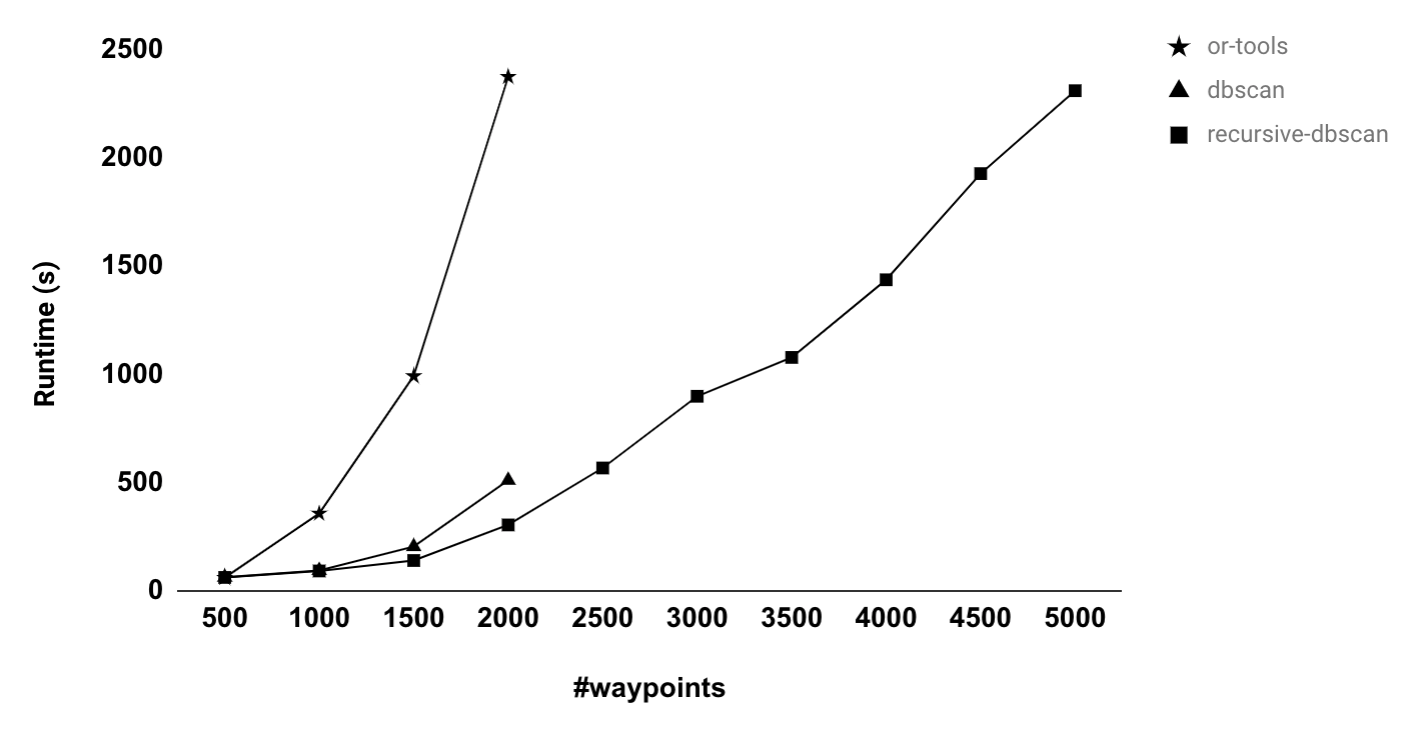}}
\caption{Average runtime for each solution.} \label{results-runtime}
\end{figure}

As can be reasoned from Fig. 5., the runtimes are exponentially increasing. The Recursive-DBSCAN curve possesses the lowest gradient, followed by DBSCAN. The OR-tools solution has got the greatest gradient.

\subsection{Overall Performance of the Algorithms}

\begin{table}[H]
\caption{Average percentage difference between OR-tools and other methods}
\centering
\begin{tabular}{l c c c} \toprule
Metric & \text{OR-tools} & \text{Recursive-DBSCAN} & DBSCAN \\ \midrule
Total Distance & \textbf{0.0\%} & +3.6\% &  +120\% \\
Runtime & 0.0\% & \textbf{-61\%} & -57\% \\
\#busy vehicles &\textbf{ 0.0\%} & +6.4\% & +150\% \\
\bottomrule
\end{tabular}
\label{RO-results-summary}
\end{table}

Both Recursive-DBSCAN and DBSCAN methods lead to a significant drop in runtime. DBSCAN has large increase in total distance travelled and the number of busy vehicle as measured against OR-tools. For Recursive-DBSCAN, the differences in total distance travelled and the number of busy vehicles are below 3.6\% and 6.4\% respectively. Summary fo results is presented in Table 3.

\section{Discussion}

As we can see in Fig. 3, Fig. 4, Fig. 5 and Table 3, our proposed Recursive-DBSCAN method has better runtimes than both the DBSCAN and OR-tools (by $61\%$) methods. It is because Recursive-DBSCAN divides the node-set into a lower number of clusters, which decreases the number of times the OR-tools Routing library needs to be invoked. Both Recursive-DBSCAN and DBSCAN have better runtimes than OR-tools as they do not have to invoke OR-tools Routing library for more than max\_cluster\_size  number of nodes.

The total distance for Recursive-DBSCAN is lower than for DBSCAN and $3.6\%$ higher than for OR-tools.  In terms of total distance, Recursive-DBSCAN performs better than DBSCAN because it creates less clusters, which in turn means that less vehicles are required, which leads to lower distance. However, Recursive-DBSCAN performs worse than OR-tools, because it does not allow vehicles to be assigned to orders within different clusters and therefore does not take into account the relationships between the clusters.

Both the DBSCAN and the OR-tools solvers would crash at 2000 waypoints due to an extreme RAM usage, while the Recursive-DBSCAN method would find solutions up to 5000 waypoints.

In summary, Recursive-DBSCAN performs better than OR-tools in terms of runtime thanks to dividing the node-set into smaller clusters, for which the optimisation results can be obtained more quickly than for the whole set being optimised at once, as the number of possible assignments and routes is greatly limited by breaking down the problem into constituent parts. Having less possible solutions to be checked means that the optimisation finishes more quickly.

At the same time, Recursive-DBSCAN performs better than DBSCAN in terms of total distance and number of vehicles used, as it simply creates less clusters than DBSCAN, which also leads to clusters containing more nodes on average. That allows the optimisation within clusters to be of better quality, as there is enough nodes to fully utilise the capacity of each vehicle.

Most importantly, the Recursive-DBSCAN method runtimes are more acceptable to be used in the `on-demand' economy, as it reduces the time drivers would be required to wait compared to OR-tools and supports instances of more than 2000 nodes.

\section{Conclusion}
This paper has described Recursive-DBSCAN - a new clustering approach to solve Capacitated Vehicle Routing Problem. By recursively clustering nodes together in batches that contain a similar number of nodes, we were able to reduce the runtimes by $61\%$ as benchmarked against Google Optimization Tools. That improvement is caused by splitting the node-set into constituent parts, consequently reducing the possible number of assignments and routes. At the same time, the total distance for Recursive-DSBCAN routes was $3.6\%$ higher than for OR-tools. Currently, the presented approach is used by GOGOVAN to optimise their routes on a daily basis.

We look forward to continue the work in that field. The next steps are to implement and benchmark the Recursive-DBSCAN clustering approach against some state-of-the-art solvers \citep{Uchoa16} using the test dataset provided by \citep{Uchoa16} as well. We hope to observe an improvement in runtimes against the state-of-the-art approaches, similarly to the improvement we have observed against Google Optimization Tools routing solver.

\bigskip

\appendix

\section{Testing Results}

Below, we present the exact results of each CVRPTW optimisation request we have run. For data privacy reasons, unfortunately we cannot share the exact input.

\begin{table}[H]
\caption{Results}
\centering 
\hspace*{-1cm}
\begin{tabular}{lccccccccc} \toprule
\#wps & runtime & runtime & runtime & distance & distance & distance & no\_cars & no\_cars & no\_cars \\
 & or-tools & dbscan & recursive & or-tools & dbscan & recursive & or-tools & dbscan & recursive \\
500 & 60 & 62 & 65 & 483388 & 516713 & 515871 & 17 & 18 & 18 \\
500 & 72 & 65 & 75 & 486747 & 507389 & 511661 & 17 & 18 & 18 \\
500 & 60 & 61 & 64 & 486239 & 515871 & 509121 & 17 & 18 & 18 \\
500 & 63 & 67 & 66 & 485711 & 500271 & 503640 & 17 & 18 & 18 \\
500 & 70 & 72 & 60 & 485591 & 499130 & 498963 & 17 & 18 & 18 \\
500 & 61 & 70 & 60 & 485347 & 498963 & 498963 & 17 & 18 & 18 \\
500 & 54 & 63 & 58 & 462337 & 469396 & 468921 & 17 & 18 & 18 \\
500 & 59 & 61 & 64 & 462812 & 470426 & 470782 & 17 & 18 & 18 \\
500 & 59 & 61 & 59 & 464018 & 470782 & 468593 & 17 & 18 & 18 \\
500 & 67 & 61 & 62 & 505695 & 477230 & 477230 & 17 & 18 & 18 \\
500 & 56 & 63 & 61 & 505014 & 477985 & 477230 & 17 & 18 & 18 \\
500 & 58 & 60 & 60 & 504797 & 477230 & 478540 & 17 & 18 & 18 \\
500 & 59 & 60 & 59 & 489643 & 536242 & 532264 & 17 & 18 & 18 \\
500 & 58 & 61 & 59 & 489643 & 535311 & 535799 & 17 & 18 & 18 \\
500 & 58 & 59 & 59 & 489671 & 535311 & 535311 & 17 & 18 & 18 \\
1000 & 346 & 74 & 74 & 740345 & 1264443 & 767720 & 34 & 71 & 38 \\
1000 & 351 & 86 & 83 & 740738 & 1264306 & 765552 & 34 & 71 & 38 \\
1000 & 290 & 78 & 70 & 759366 & 1267402 & 766308 & 34 & 71 & 38 \\
1000 & 295 & 67 & 72 & 725851 & 1460247 & 720003 & 34 & 89 & 37 \\
1000 & 316 & 70 & 85 & 724167 & 1460504 & 719268 & 34 & 89 & 37 \\
1000 & 346 & 74 & 75 & 725053 & 1460075 & 720048 & 34 & 89 & 37 \\
1000 & 402 & 136 & 139 & 753055 & 1251045 & 742976 & 34 & 72 & 36 \\
1000 & 361 & 118 & 112 & 753055 & 1257972 & 747099 & 34 & 72 & 36 \\
1000 & 483 & 148 & 137 & 747869 & 1256926 & 744831 & 34 & 72 & 36 \\
1000 & 482 & 122 & 114 & 714253 & 1650897 & 748432 & 34 & 97 & 37 \\
1000 & 489 & 121 & 113 & 714245 & 1650976 & 750093 & 34 & 97 & 37 \\
1000 & 353 & 125 & 111 & 719354 & 1651073 & 748446 & 34 & 97 & 37 \\
1000 & 288 & 65 & 65 & 715072 & 1360899 & 694867 & 34 & 81 & 37 \\
1000 & 280 & 65 & 65 & 715170 & 1360899 & 694407 & 34 & 81 & 37 \\
1000 & 281 & 65 & 65 & 715170 & 1360899 & 694407 & 34 & 81 & 37 \\
1500 & 943 & 201 & 136 & 922991 & 3175289 & 943656 & 51 & 198 & 54 \\
1500 & 957 & 202 & 135 & 923117 & 3175312 & 943624 & 51 & 198 & 54 \\
1500 & 1118 & 202 & 135 & 924407 & 3175366 & 943618 & 51 & 198 & 54 \\
1500 & 965 & 254 & 156 & 927529 & 2845066 & 955311 & 51 & 193 & 54 \\
1500 & 948 & 195 & 135 & 927529 & 2840604 & 955003 & 51 & 193 & 54 \\
1500 & 946 & 195 & 135 & 927529 & 2840025 & 955003 & 51 & 193 & 54 \\
1500 & 946 & 187 & 124 & 901753 & 2509369 & 945549 & 51 & 152 & 54 \\
1500 & 942 & 188 & 126 & 901679 & 2509369 & 945549 & 51 & 152 & 54 \\
1500 & 952 & 189 & 124 & 901753 & 2509423 & 945549 & 51 & 152 & 54 \\
1500 & 948 & 202 & 146 & 920773 & 2974830 & 959563 & 51 & 194 & 54 \\
1500 & 954 & 203 & 147 & 920773 & 2976673 & 959563 & 51 & 194 & 54 \\
1500 & 944 & 205 & 146 & 920773 & 2976102 & 959563 & 51 & 194 & 54 \\
1500 & 953 & 197 & 144 & 954255 & 2691762 & 1006466 & 51 & 171 & 55 \\
1500 & 1130 & 196 & 145 & 948436 & 2692423 & 1006466 & 51 & 171 & 55 \\
1500 & 1222 & 249 & 171 & 949135 & 2692646 & 1005565 & 51 & 171 & 55 \\
 \bottomrule
\end{tabular}
\hspace*{-1cm}
\label{results-all-1}
\end{table}

\begin{table}[H]
\caption{Results}
\centering 
\hspace*{-1cm}
\begin{tabular}{lccccccccc} \toprule
\#wps & runtime & runtime & runtime & distance & distance & distance & no\_cars & no\_cars & no\_cars \\
 & or-tools & dbscan & recursive & or-tools & dbscan & recursive & or-tools & dbscan & recursive \\
2000 & 2738 & 625 & 371 & 1107218 & 3202771 & 1179192 & 67 & 198 & 69 \\
2000 & 2642 & 598 & 344 & 1106752 & 3203236 & 1175043 & 67 & 198 & 69 \\
2000 & 2603 & 643 & 358 & 1105813 & 3203673 & 1176404 & 67 & 198 & 69 \\
2000 & 2392 & 606 & 292 & 1085628 & 3556857 & 1171007 & 67 & 239 & 72 \\
2000 & 2415 & 532 & 297 & 1084229 & 3560283 & 1171322 & 67 & 239 & 72 \\
2000 & 2606 & 497 & 289 & 1084432 & 3559491 & 1171157 & 67 & 239 & 72 \\
2000 & 2415 & 572 & 336 & 1081554 & 3031531 & 1215245 & 67 & 201 & 70 \\
2000 & 2210 & 451 & 298 & 1081554 & 3033713 & 1213129 & 67 & 201 & 70 \\
2000 & 2237 & 452 & 301 & 1081554 & 3034193 & 1213188 & 67 & 201 & 70 \\
2000 & 2216 & 452 & 290 & 1071946 & 2919687 & 1125675 & 67 & 178 & 70 \\
2000 & 2216 & 450 & 290 & 1071846 & 2921376 & 1123879 & 67 & 178 & 70 \\
2000 & 2225 & 448 & 289 & 1071946 & 2920938 & 1123718 & 67 & 178 & 70 \\
2000 & 2222 & 445 & 271 & 1081664 & 2956164 & 1127928 & 67 & 186 & 70 \\
2000 & 2226 & 442 & 270 & 1081715 & 2956177 & 1127850 & 67 & 186 & 70 \\
2000 & 2220 & 443 & 270 & 1081664 & 2956173 & 1131021 & 67 & 186 & 70 \\
2500 & - & - & 527 & - & - & 1358060 & - & - & 89 \\
2500 & - & - & 563 & - & - & 1355550 & - & - & 89 \\
2500 & - & - & 539 & - & - & 1358200 & - & - & 89 \\
2500 & - & - & 547 & - & - & 1390752 & - & - & 89 \\
2500 & - & - & 556 & - & - & 1390425 & - & - & 89 \\
2500 & - & - & 520 & - & - & 1389660 & - & - & 89 \\
2500 & - & - & 603 & - & - & 1440061 & - & - & 86 \\
2500 & - & - & 565 & - & - & 1439611 & - & - & 86 \\
2500 & - & - & 578 & - & - & 1450749 & - & - & 86 \\
2500 & - & - & 558 & - & - & 1412417 & - & - & 88 \\
2500 & - & - & 581 & - & - & 1411956 & - & - & 88 \\
2500 & - & - & 675 & - & - & 1421848 & - & - & 88 \\
2500 & - & - & 611 & - & - & 1455259 & - & - & 89 \\
2500 & - & - & 561 & - & - & 1455195 & - & - & 89 \\
2500 & - & - & 524 & - & - & 1453103 & - & - & 89 \\
3000 & - & - & 751 & - & - & 1633968 & - & - & 107 \\
3000 & - & - & 844 & - & - & 1625904 & - & - & 107 \\
3000 & - & - & 946 & - & - & 1631154 & - & - & 107 \\
3000 & - & - & 890 & - & - & 1665061 & - & - & 107 \\
3000 & - & - & 949 & - & - & 1666696 & - & - & 107 \\
3000 & - & - & 953 & - & - & 1665962 & - & - & 107 \\
3000 & - & - & 1061 & - & - & 1600832 & - & - & 105 \\
3000 & - & - & 1077 & - & - & 1606483 & - & - & 105 \\
3000 & - & - & 947 & - & - & 1605001 & - & - & 105 \\
3000 & - & - & 938 & - & - & 1608125 & - & - & 105 \\
3000 & - & - & 932 & - & - & 1605236 & - & - & 105 \\
3000 & - & - & 900 & - & - & 1608508 & - & - & 105 \\
3000 & - & - & 784 & - & - & 1673911 & - & - & 106 \\
3000 & - & - & 746 & - & - & 1674320 & - & - & 106 \\
3000 & - & - & 752 & - & - & 1673882 & - & - & 106 \\
 \bottomrule
\end{tabular}
\hspace*{-1cm}
\label{results-all-2}
\end{table}

\begin{table}[H]
\caption{Results}
\centering 
\hspace*{-1cm}
\begin{tabular}{lccccccccc} \toprule
\#wps & runtime & runtime & runtime & distance & distance & distance & no\_cars & no\_cars & no\_cars \\
 & or-tools & dbscan & recursive & or-tools & dbscan & recursive & or-tools & dbscan & recursive \\
3500 & - & - & 1153 & - & - & 1860161 & - & - & 123 \\
3500 & - & - & 1102 & - & - & 1869332 & - & - & 123 \\
3500 & - & - & 1081 & - & - & 1868842 & - & - & 123 \\
3500 & - & - & 1052 & - & - & 1813835 & - & - & 124 \\
3500 & - & - & 1047 & - & - & 1813956 & - & - & 124 \\
3500 & - & - & 1048 & - & - & 1816695 & - & - & 124 \\
3500 & - & - & 1105 & - & - & 1850848 & - & - & 125 \\
3500 & - & - & 1106 & - & - & 1850695 & - & - & 125 \\
3500 & - & - & 1106 & - & - & 1850796 & - & - & 125 \\
3500 & - & - & 1040 & - & - & 1815906 & - & - & 124 \\
3500 & - & - & 1040 & - & - & 1816712 & - & - & 124 \\
3500 & - & - & 1040 & - & - & 1815959 & - & - & 124 \\
3500 & - & - & 1080 & - & - & 1823448 & - & - & 122 \\
3500 & - & - & 1083 & - & - & 1824033 & - & - & 122 \\
3500 & - & - & 1078 & - & - & 1823337 & - & - & 122 \\
4000 & - & - & 1292 & - & - & 1999660 & - & - & 145 \\
4000 & - & - & 1281 & - & - & 1995747 & - & - & 145 \\
4000 & - & - & 1283 & - & - & 1995751 & - & - & 145 \\
4000 & - & - & 1315 & - & - & 1960920 & - & - & 142 \\
4000 & - & - & 1326 & - & - & 1961180 & - & - & 142 \\
4000 & - & - & 1423 & - & - & 1963337 & - & - & 142 \\
4000 & - & - & 1398 & - & - & 1991476 & - & - & 142 \\
4000 & - & - & 1344 & - & - & 1995379 & - & - & 142 \\
4000 & - & - & 1401 & - & - & 1991310 & - & - & 142 \\
4000 & - & - & 1383 & - & - & 2037947 & - & - & 141 \\
4000 & - & - & 1758 & - & - & 2038129 & - & - & 141 \\
4000 & - & - & 1446 & - & - & 2035300 & - & - & 141 \\
4000 & - & - & 1600 & - & - & 1966531 & - & - & 140 \\
4000 & - & - & 1774 & - & - & 1967573 & - & - & 140 \\
4000 & - & - & 1510 & - & - & 1962472 & - & - & 140 \\
4500 & - & - & 1828 & - & - & 2190689 & - & - & 160 \\
4500 & - & - & 1820 & - & - & 2192041 & - & - & 160 \\
4500 & - & - & 1828 & - & - & 2191594 & - & - & 160 \\
4500 & - & - & 1601 & - & - & 2149448 & - & - & 160 \\
4500 & - & - & 1603 & - & - & 2149016 & - & - & 160 \\
4500 & - & - & 1613 & - & - & 2150206 & - & - & 160 \\
4500 & - & - & 2006 & - & - & 2212566 & - & - & 160 \\
4500 & - & - & 2244 & - & - & 2214822 & - & - & 160 \\
4500 & - & - & 2040 & - & - & 2208680 & - & - & 160 \\
4500 & - & - & 1901 & - & - & 2198181 & - & - & 159 \\
4500 & - & - & 1993 & - & - & 2195011 & - & - & 159 \\
4500 & - & - & 2234 & - & - & 2194019 & - & - & 159 \\
4500 & - & - & 1986 & - & - & 2227280 & - & - & 161 \\
4500 & - & - & 1837 & - & - & 2225391 & - & - & 161 \\
4500 & - & - & 2348 & - & - & 2233762 & - & - & 161 \\
 \bottomrule
\end{tabular}
\hspace*{-1cm}
\label{results-all-3}
\end{table}

\begin{table}[H]
\caption{Results}
\centering 
\hspace*{-1cm}
\begin{tabular}{lccccccccc} \toprule
\#wps & runtime & runtime & runtime & distance & distance & distance & no\_cars & no\_cars & no\_cars \\
 & or-tools & dbscan & recursive & or-tools & dbscan & recursive & or-tools & dbscan & recursive \\
 5000 & - & - & 3127 & - & - & 2453175 & - & - & 179 \\
5000 & - & - & 2754 & - & - & 2446140 & - & - & 179 \\
5000 & - & - & 2458 & - & - & 2449849 & - & - & 179 \\
5000 & - & - & 2148 & - & - & 2416567 & - & - & 178 \\
5000 & - & - & 2230 & - & - & 2417642 & - & - & 178 \\
5000 & - & - & 2235 & - & - & 2417174 & - & - & 178 \\
5000 & - & - & 2150 & - & - & 2424674 & - & - & 179 \\
5000 & - & - & 2094 & - & - & 2424928 & - & - & 179 \\
5000 & - & - & 2081 & - & - & 2425004 & - & - & 179 \\
5000 & - & - & 2328 & - & - & 2419737 & - & - & 181 \\
5000 & - & - & 2340 & - & - & 2419498 & - & - & 181 \\
5000 & - & - & 2325 & - & - & 2419580 & - & - & 181 \\
5000 & - & - & 2112 & - & - & 2380606 & - & - & 180 \\
5000 & - & - & 2122 & - & - & 2379298 & - & - & 180 \\
5000 & - & - & 2116 & - & - & 2379444 & - & - & 180 \\
 \bottomrule
\end{tabular}
\hspace*{-1cm}
\label{results-all-4}
\end{table}

\end{document}